# FuDFEND: Fuzzy-domain for Multi-domain Fake News Detection

Chaoqi Liang, Yu Zhang*, Xinyuan Li, Jinyu Zhang and Yongqi Yu

Harbin Institute of Technology, Harbin, China
{120L030706,7203610518,7203610523,7203610515}@stu.hit.edu.cn
zhangyu@ir.hit.edu.cn

**Abstract.** On the Internet, fake news exists in various domain (e.g., education, health). Since news in different domains has different features, researchers have begun to use single domain label for fake news detection recently. Existing works show that using single domain label can improve the accuracy of fake news detection model. However, there are two problems in previous works. Firstly, they ignore that a piece of news may have features from different domains. The single domain label focuses only on the features of one domain. This may reduce the performance of the model. Secondly, their model cannot transfer the domain knowledge to the other dataset without domain label. In this paper, we propose a novel model, FuDFEND[1], which solves the limitations above by introducing the fuzzy inference mechanism. Specifically, FuDFEND utilizes a neural network to fit the fuzzy inference process which constructs a fuzzy domain label for each news item. Then, the feature extraction module uses the fuzzy domain label to extract the multi-domain features of the news and obtain the total feature representation. Finally, the discriminator module uses the total feature representation to discriminate whether the news item is fake news. The results on the Weibo21 show that our model works better than the model using only single domain label. In addition, our model transfers domain knowledge better to Thu dataset which has no domain label.

**Keywords:** Fake News Detection, Social Media, Multi-domain.

## 1 Introduction

With the development of the Internet, social media platforms such as Sina Weibo and Twitter have become the main source of information. Fake news is also widely spread on social platforms. The fake news on social media is usually a short text, a picture and a short video that they can be understood in a few seconds. At the same time, the fake news incites people's emotions and stimulates users to forward, so they can be widely spread. Tavernise [1] showed that important social events were affected due to moderated fake news campaigns. The spread of fake news may result in people's

---

panic and social dislocation. The quality of content on social media platforms has suffered greatly due to the spread of fake news, misinformation and unverifiable facts [2]. Therefore, it is meaningful to research into social media fake news detection today.

At present, there are two methods of social media fake news detection. One is social background detection method, and the other is content-based detection method [3]. Social background method aims to study users' social network structure, users' personal information, microblog forwarding and reply relationship and rumor's propagation patterns. The content-based research method aims to detect the text, voice and video carried by microblog news. For example, Mouratidis et al. [4] analyzes linguistic features to distinguish between fake news and real news. Our work focuses on the content-based method.

On the Internet, fake news arises in many different domains. The task of detecting fake news in multiple domains is called Multi-domain Fake News Detection (MFND). There are two main challenges in MFND. First, the performance of techniques generally drops if news records are coming from different domains (e.g., politics, entertainment), and a possible explanation of this could be the rather unique content and style of each domain [5]. For example, during the 2020 US election, political fake news and Covid19 fake news were widespread simultaneously. It is difficult to detect both political fake news and Covid19 fake news at the same time, because of the different or even conflicting features between political fake news and Covid19 news. Thus, a variety of previous works [6,7, 8, 9, 10] focused on rumor detection in a single domain. But herein lies another problem. In a single-domain, there may be too little data to train a good model. Therefore, Nan et al. [11] proposed a **M**ulti-**d**omain **F**ake **N**ews **D**etection Model (MDFEND) which use single domain label to solve the above two problems. Single domain label is to describe that a piece of news belongs to a certain domain, such as science and technology, education, health and so on. The model receives news content and single domain label as input. Utilizing these data, the model can extract the common features of fake news in all different domains, and can distinguish the specific features of certain domain by domain label.

**Table 1.** The difference between single domain label and fuzzy domain label.

| News content | Single domain label | Fuzzy domain label |
|---|---|---|
| The new generation of iphone has added numbers of new technologies such as face recognition, which are popular with consumers. The next day, Apple shares rose sharply. | Technology | Technology: 70% Business: 20% Finance: 10% |

We build this paper on that work [11] by recognizing that there are two problems. (i) A piece of news may have features of several domains. There is an example showed in Table 1. This news has the features from three domains. Single domain label can't help the model extract this kind of news with multi domain features. (ii) Their model can't be transferred to datasets without domain label. Therefore, we propose a novel model, **Fu**zzy-**d**omain **F**ake **N**ews **D**etection Model (FuDFEND). FuDFEND is an improved model based on the MDFEND proposed in [11]. The improvement is made by introducing a fuzzy inference mechanism into the model. The fuzzy inference mechanism can solve the above two problems. The fuzzy inference mechanism constructs a fuzzy domain label for each news item. Compared with single domain label, fuzzy domain label can better describe the domain features of news, so that help the model better extract the multi-domain features of news. We demonstrated this by the experiment on the weibo21. In addition, FuDFEND has better transfer ability than MDFEND. We will illustrate this through experiment on Thu dataset.

Our summarized contributions with this paper are:

- We propose a novel model, Fuzzy-domain Fake News Detection Model (FuDFEND), which can extract the multi-domain features of news content by the fuzzy domain label that fuzzy inference mechanism generate. The results on the Weibo21 show that FuDFEND works better than the model, MDFEND, using only single domain label.
- In order to describe the multi-domain features of news, we introduced fuzzy inference mechanism to the multi-domain fake news detection task. The fuzzy domain label constructed by fuzzy inference mechanism can more accurately describe the multi-domain features that news has.
- We solve the problem that model can't transfer to the dataset without domain label. We evaluate FuDFEND on Thu dataset which has no domain label. Experimental results show that FuDFEND can transfer domain knowledge better by utilizing the fuzzy inference mechanism.

## 2 Related Work

### 2.1 Fake news Detection Methods

Researchers have come up with number of ways to detect fake news.

**Content-based Methods.** In [12], the authors proposed an ensemble classification model for fake news detection. Their model obtains relevant features from a fake news dataset and then uses an ensemble model to classify the extracted features, but these works are based only on news texts. In [13], the authors study and analyze whether fake news can be distinguished from mainstream news by text writing style and whether fake news can be detected only by writing style by building a model. Rawat et al. [2] proposed a method to automatically collect fake news detection tasks online. For each piece of news data, they collect evidence and generate their summaries as another input to the model to help detect news text. In [14] author found images have different distribution patterns and statistically distinctive patterns for fake and real news. It reveals that images in the real news are more diverse and much denser than those in the fake news. Therefore, they extract image features from the visual features and overall statistics of images in news events. Alonso-Bartolome et al. [15] used the early fusion method to fuse text and image information for rumor detection. In [16], an innovative RNN with attention mechanism (att-RNN) is proposed for effective multimodal feature fusion. In [17], author propose a Multi-domain Visual Neural Network (MVNN) framework consisting of three main parts to mix all the features together.

**Social Background Detection Method.** These researches mainly focused on propagation patterns, publisher-news relations and user-news interactions. In [18], the authors examined numerous features related to user, linguistic, network and temporal characteristics of rumors. They studied the spread patterns of rumors over time and the ability to track the precise changes in the predictive power of rumor features. However, the events of this volume (111 rumor and non-romors) may not be sufficient for summarizing the prediction performance, and the algorithm takes every feature into account did not achieve good results in short observation periods. Shu et al. [19] examined the correlation of publisher bias, news stance, and relevant user engagement. They studied the novel problem of exploiting social context for fake news detection and proposed a tri-relationship embedding framework TriFN, which model relations and interactions between publishers, news, and users simultaneously for fake news classification. But it requires the social context information to be included in the data. Meanwhile, there may be significant differences in social relations in different domains, and targeted detection in different domains is a promising concept. Alrubaian et al. [20] proposed a credibility analysis system consisting of a model based on reputation, a feature ranking algorithm, a credibility evaluation classifier engine, and a user expertise model, to assess the accuracy of information on Twitter so it can stop the promotion of disinformation.

## 2.2    Multi-domain Rumor Task

In [5], the author introduced two new English fake news datasets (FakeNewsAMT, Celebrity) covering six news domains, and the author analyzes the different performance of his model in different news domains. Nan et al. [11] construct Weibo21, a multi-domain fake news dataset in Chinese. Weibo21 contains more than 9000 pieces of real and fake news from nine domains and was manually marked the domain label for each piece of news. And they designed a Multi-domain Fake News Detection Model (MDFEND) for Weibo21. The model can extract the common features of fake news in all different domains, and can distinguish the specific features of certain domain through single domain label. However, this model relies on manually annotated datasets, which not only contain news text and true and false information, but also need to input the domain category to which the news belongs, then select the corresponding vector in the domain matrix according to the type of domain to calculate the weight of each domain experts (TextCNN), so it requires a lot of effort to obtain the data set that match the model. In the following we will propose our improvement strategy on the base of this model.

## 3    FuDFEND: Fuzzy-domain Fake News Detection Model

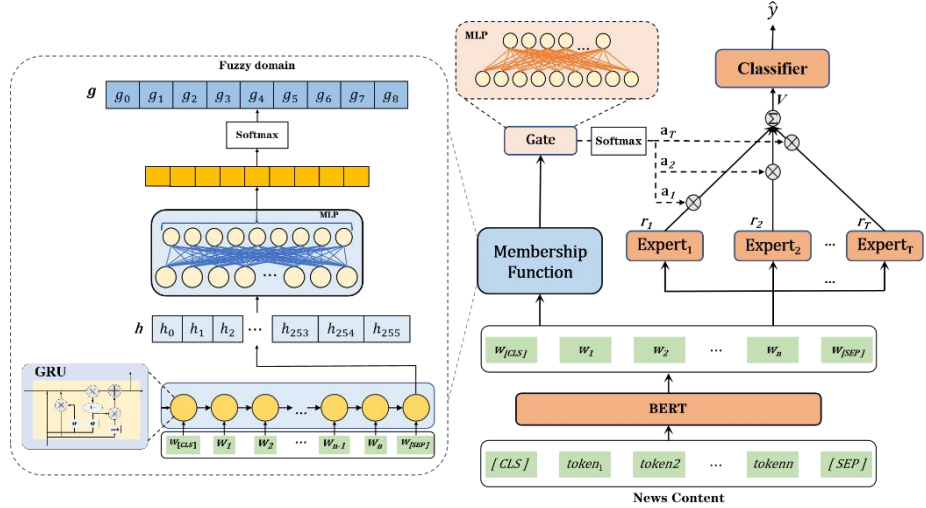

**Fig. 1.** Overall framework of FuDFEND.

FuDFEND is an improved model based on the MDFEND proposed in [11]. The improvement is made by introducing a fuzzy inference mechanism into the model. The fuzzy inference mechanism includes two modules: Membership Function and Gate.

Overall framework of FuDFEND is showed in figure 1. The working process of the model is as follows. Enter the news text into BERT [21, 22] to obtain a series of word embeddings $W=[w_{[CLS]}, w_1, w_2, ..., w_n, w_{[SEP]}]$. $W$ is input into a mixture of experts to extract the features of different domains. Enter $W$ into Membership Function to obtain fuzzy domain label. Then, Gate generates weight scores $\boldsymbol{\alpha}$ by inputting fuzzy domain label. The output experts are weighted and summed by the weight scores $\boldsymbol{\alpha}$, so that obtain the total feature representation $\boldsymbol{v}$. The classifier module uses the total feature representation $\boldsymbol{v}$ to discriminate whether the news item is fake news.

### 3.1    Membership Function

In traditional set theory, we define a set with a definite condition. For example, we define a tall set as those who are taller than or equal to 190cm. The other belong to short set. If a person is 189cm tall, very close to 190cm but still belongs to short set. Obviously, the traditional way of describing sets is very crude. In order to describe the set more accurately, LA Zadeh proposed the concept of fuzzy sets in 1965. A fuzzy set is characterized by a Membership (characteristic) Function which assigns to each object a membership grade between zero and one [23]. As shown in Figure 2, we can use a Membership Function to describe high set and short set. If a person is 189cm tall, he belongs to a high set with a membership grade of 0.9 and belongs to a short set with a membership grade of 0.1.

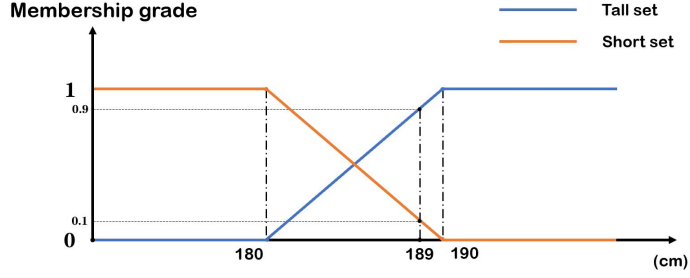

**Fig. 2.** An example of fuzzy set.

A piece of news may have features of different domains. Here is an example that "The new generation of iphone has added a number of new technologies such as face recognition, which are popular with consumers. The next day, Apple shares rose sharply." This news has the features of both science domain and financial domain. Therefore, we need to use fuzzy sets to more precisely measure the domains that news belongs to.

We will show how to use neural network to fit the Membership Function for the sets of news domains. The Membership Function consist of an *GRU*, a multi-layer perception (*MLP*) and a softmax function. Membership Function can generate a nine-dimensional membership grade vector $\boldsymbol{g}$. We call this vector $\boldsymbol{g}$ as the fuzzy domain label. During training, we train it as training a classifier for news domains. When we use it, we regard it as a Membership Function. The specific process is as follows. For a piece of news, we put the news' word embeddings $\boldsymbol{W}$ into an *GRU*. The output $\boldsymbol{h}$ of *GRU* is fed into *MLP* which output is a nine-dimensional vector. Use softmax function to normalize the output of *MLP*. Then we obtain a vector $\boldsymbol{g}$ which has nine dimensions and sum of all dimensions is 1. The value of each dimension of the vector represents the membership grade of nine news domains (Science, Military, Education, Disasters, Politics, Health, Finance, Entertainment). We denote the Membership Function as $M(\cdot\,;\,\cdot)$, and $\theta, \theta_1, \theta_2$ is the parameters in the Membership Function, *GRU*, *MLP*. $\boldsymbol{g}$ represents the fuzzy domain label:

$$\boldsymbol{g} = M(\boldsymbol{W}; \theta) = softmax(MLP(GRU(\boldsymbol{W};\, \theta_1);\, \theta_2)) \tag{1}$$

Membership Function's parameter is pre-trained. During the training of rumor detection task, the parameter of Membership Function is fixed. Membership Function is train as a as a classifier, its accuracy rate can only reach about 82% in the end. But as a Membership Function, We cannot think that its error rate is 18%, but should be understood as the correction of labels by large models. Because some domain labels labeled by the human are not necessarily suitable for machine. The machine reclassified things that it thought had the same features. Further, the outputs of Membership Function, $\boldsymbol{g}$, is a probability distribution, which can provide enough noise for the model.

Figure 3 shows an execution of the Membership Function. Membership Function analysis shows that the content of the news has the features of entertainment domain and social domain.

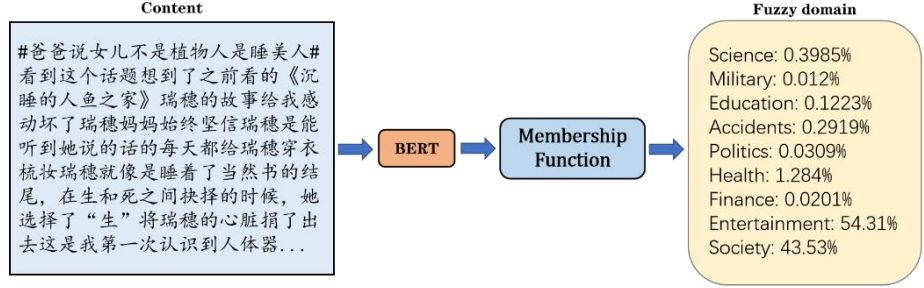

**Fig. 3.** An execution of the Membership Function. The single domain label of this news is "Society". The content of this news is that "# Dad said his daughter is not a vegetative person but a sleeping beauty # when I saw this topic, I was moved by the story of Mizuho in *The House Where the Mermaid Sleeps*. Mizuho's mother always believed that Mizuho could hear what she said. She dressed and dressed Mizuho every day. Mizuho was like falling asleep. Of course, at the end of the book, when choosing between life and death, she chose "living" to donate Mizuho's heart. This is the first time I realized the human body."

### 3.2 Feature Extraction

We use Mixture-of-Expert [24, 25, 26] to extract features of news' content. Each expert network is a TextCNN [27] in our model. Each expert has its own area of expertise and is good at extracting the features of certain domain. An expert network can be denoted by $E_i(\cdot\ ;\ \cdot)\,(1 \leq i \leq T)$. $T$ is a hyperparameter, representing the number of the experts. $\boldsymbol{r}_i$ denote the output of the expert. $\varphi_i$ are the trainable parameters of an expert:

$$\boldsymbol{r}_i = E_i(\boldsymbol{W};\ \varphi_i) = TextCNN_i(\boldsymbol{W};\ \varphi_i) \tag{2}$$

### 3.3 Domain Gate

Because different experts can extract the features of different domains, we input the fuzzy domain label into the Domain Gate, to obtain the weight score. The weight score consists of T positive real numbers, corresponding to T experts. Then we use the weight score to aggregate the features extracted by experts. We denote the Domain Gate as $Gate(\cdot\ ;\ \cdot)$. $\psi$ is the trainable parameters of the Domain Gate. $\boldsymbol{\alpha}$ represent the weight score:

$$\boldsymbol{\alpha} = (\alpha_1, \alpha_2, ..., \alpha_T) = Gate(\boldsymbol{g};\ \psi) = softmax(MLP(\boldsymbol{g};\ \psi)) \tag{3}$$

### 3.4 Fake News Prediction and Loss Function

Through the weight score, we aggregate the features extracted by experts to obtain the total features vector $\boldsymbol{v}$:

$$\boldsymbol{v} = \sum_{i=1}^{T} \alpha_i \boldsymbol{r}_i \tag{4}$$

Input the total features vector into a binary classifier which is MLP with a sigmoid output layer. And then we get our model's predicted value, $\hat{y}$. The value $\hat{y}$ is between zero and one and it is the probability that the news is fake. The larger the value of $\hat{y}$ is, the more likely the model is to conclude that the news is fake:

$$\hat{y} = sigmod(MLP(\boldsymbol{v}, \xi)) \tag{5}$$

$\xi$ represent the trainable parameters of the MLP.

We use $y_i$ to represent value label and $\hat{y}_i$ to represent predicted value. We employ Binary Cross-Entropy Loss for training:

$$L = -\sum_{i=1}^{N} (y_i log \hat{y}_i + (1 - y_i) log(1 - \hat{y}_i)) \tag{6}$$

## 4 Experiment

FuDFEND is an improved model based on the MDFEND proposed in [11]. The improvement is made by introducing a fuzzy inference mechanism into the model. In order to show the effect of fuzzy inference mechanism, we compare our model FuDFEND with baseline method which mentioned in [11], especially MDFEND which hasn't fuzzy inference mechanism.

### 4.1 Dataset

In this section, we introduce the two datasets we use.

Weibo21 was released in this work [11] last year. It is a Chinese multi-domain fake news dataset. It consists of 4,488 pieces of fake news and 4,640 pieces of real news, covering nine different news domains: **Science**, **Military**, **Education**, **Accidents**, **Politics**, **Health**, **Finance**, **Entertainment**. The number of true and fake news contained in each domain is shown in Table 2.

**Table 2.** Data Statistics of Weibo21.

| domain | Science | Military | Education | Accidents | Politics |
|--------|---------|----------|-----------|-----------|----------|
| real | 143 | 121 | 243 | 185 | 306 |
| fake | 93 | 222 | 248 | 591 | 546 |
| all | 236 | 343 | 491 | 776 | 852 |
| domain | Health | Finance | Entertainment | Society | All |
| real | 485 | 959 | 1000 | 1198 | 4640 |
| fake | 515 | 362 | 440 | 1471 | 4488 |

| all | 1000 | 1321 | 1440 | 2669 | 9128 |
|---|---|---|---|---|---|

Each piece of news is labeled a specific domain. We show a few examples of the Weibo21 dataset in Table 3.

**Table 3.** Five Samples of Weibo21.

| content | Single domain label | Fake label |
|---|---|---|
| 【三星......年推出柔性折叠屏的产品，你会入手么？ | Science | 0 |
| #四川......化工厂泄漏#紧急撤离！！大家注意安全 | Accidents | 1 |
| #广州......料公布！居然是日本间谍！】@广东热搜 | Politics | 1 |
| 《哈利......面都是回忆[悲伤]http://t.cn/RURkC7V | Entertainment | 0 |
| #人民......。求@平安北京@通州警方在线给说法！！ | Society | 1 |

Another dataset we used is the Chinese Rumor Dataset released by thunlp [28, 29]. The news in this dataset has no domain label. The dataset consists of two parts. One is called rumors_V170613 [28], including 31669 rumors. Another named CEO_Dataset [29], including 1538 rumors and 1849 real news. We found that some rumors in CEO_Dataset coincide with the data in Weibo21, so we randomly selected 1560 pieces of fake news from rumors_V170613 and combined the fake news with the real news in the CEO_Dataset to form a new dataset that we called Thu dataset. We performed zero-shot learning on Thu dataset to test the model's capability of domain knowledge transfer.

## 4.2    Experiment Setting

BERT [21, 22] is fixed during training. The dimension of BERT's output embedding vectors is fixed to 768. The max length of the sentence is 170. We employ the Adam [30] optimizer for our training. The learning rate is 5e-4. In order to enhance the credibility of our experiments, the process is performed for 5 times and the average f1-score is reported.

## 4.3    Train Membership Function and FuDFEND

As mentioned previously, we trained the Membership Function as a nine-classification classifier for domain label. We randomly selected 70% of the data from Weibo21 as the training set and the remain as the verification set. The mini-batch size is 32. After training 7 epochs using the crossentropy loss function, the f1-score on the validation set reaches 82.31%. We take the Membership Function from the 7th epoch training, because it has the highest f1-score. As training FuDFEND, the Membership Function is fixed. The mini-batch size is 64.

### 4.4 Experiment on Weibo21

Nan et al. [11] have showed the effectiveness of their multi-domain fake news detection model (MDFEND). Further, the results in Table 4 support that our idea is right. The fake news detection performance of FuDFEND compared with MDFEND demonstrates that the fuzzy domain label generated by fuzzy inference mechanism can better help the model extract the features for fake news detection task.

**Table 4.** Multi-domain Fake News Detection Performance on Weibo21 (F1-score).

| model | Science | Military | Education | Accidents | Politics |
|---|---|---|---|---|---|
| TextCNN_single | 0.7470 | 0.7780 | 0.8882 | 0.8310 | 0.8694 |
| BiGRU_single | 0.4876 | 0.7169 | 0.7067 | 0.7625 | 0.8477 |
| BERT_single | 0.8192 | 0.7795 | 0.8136 | 0.7885 | 0.8188 |
| TextCNN_all | 0.7254 | 0.8839 | 0.8362 | 0.8222 | 0.8561 |
| BiGRU_all | 0.7269 | 0.8724 | 0.8138 | 0.7935 | 0.8356 |
| BERT_all | 0.7777 | 0.9072 | 0.8331 | 0.8512 | 0.8366 |
| EANN | 0.8225 | 0.9274 | 0.8624 | 0.8666 | 0.8705 |
| MMOE | **0.8755** | 0.9112 | 0.8706 | 0.8770 | 0.8620 |
| MOSE | 0.8502 | 0.8858 | 0.8815 | 0.8672 | 0.8808 |
| EDDFN | 0.8186 | 0.9137 | 0.8676 | 0.8786 | 0.8478 |
| MDFEND | 0.8301 | 0.9389 | **0.8917** | 0.9003 | 0.8865 |
| FuDFEND (our) | 0.8133 | **0.9468** | **0.8917** | **0.9059** | **0.9013** |
| model | Health | Finance | Entertainment | Society | All |
| TextCNN_single | 0.9053 | 0.7909 | 0.8591 | 0.8727 | 0.8380 |
| BiGRU_single | 0.8378 | 0.8109 | 0.8308 | 0.6067 | 0.7342 |
| BERT_single | 0.8909 | 0.8464 | 0.8638 | 0.8242 | 0.8272 |
| TextCNN_all | 0.8768 | 0.8638 | 0.8456 | 0.8540 | 0.8686 |
| BiGRU_all | 0.8868 | 0.8291 | 0.8629 | 0.8485 | 0.8595 |
| BERT_all | 0.9090 | 0.8735 | 0.8769 | 0.8577 | 0.8795 |
| EANN | 0.9150 | 0.8710 | 0.8957 | 0.8877 | 0.8975 |
| MMOE | 0.9364 | 0.8567 | 0.8886 | 0.8750 | 0.8947 |
| MOSE | 0.9179 | 0.8672 | 0.8913 | 0.8729 | 0.8939 |
| EDDFN | 0.9379 | 0.8636 | 0.8832 | 0.8689 | 0.8919 |
| MDFEND | 0.9400 | **0.8951** | 0.9066 | 0.8980 | 0.9137 |
| FuDFEND (our) | **0.9417** | 0.8901 | **0.9161** | **0.9174** | **0.9213** |

### 4.5 Experiment on Thu dataset

The news in Thu dataset hasn't domain label. FuDFEND has Membership Function module, so it only needs to enter news content and does not need to enter domain label. To demonstrate the transfer learning capabilities of our models, we use MDFEND as a baseline and used the hyperparameters provided in [11] to train MDFEND. MDFEND requires domain label, so we use Membership Function to label the single domain label for the dataset. The results are showed in Table 5. The results demonstrate that FuDFEND has good transfer ability with the help of fuzzy inference mechanism.

**Table 5.** Multi-domain Fake News Detection Performance on Thu dataset (F1-score).

| model | Science | Military | Education | Accidents | Politics |
|---|---|---|---|---|---|
| MDFEND | 0.8157 | **0.8407** | 0.7995 | 0.6878 | 0.7938 |
| FuDFEND (our) | **0.8760** | 0.8301 | **0.8568** | **0.7129** | **0.8148** |

| model | Health | Finance | Entertainment | Society | All |
|---|---|---|---|---|---|
| MDFEND | 0.8364 | 0.7535 | 0.8047 | 0.8642 | 0.8731 |
| FuDFEND (our) | **0.8753** | **0.7930** | **0.8492** | **0.8653** | **0.8900** |

## 5 Conclusion

Previous work has demonstrated that the use of single domain label can effectively improve the performance of fake news detection models. However, for news containing multi-domain features, models using single domain label do not synthesize multi-domain features well. In this work, we propose FuDFEND and provide a set of experiments to demonstrate that: (i) Fuzzy domain label more accurately portrays the multi-domain features of news content, so it can help the model to better extract the multi-domain features of news. (ii) Fuzzy inference mechanisms can be very helpful for model to transfer domain knowledge to dataset without domain label.

## 6 Future Work

We have received valuable comments from reviewers. For example, analyze good and bad cases and test more datasets and indicators. These suggestions give us great inspiration. However, due to time constraints and limited space, we cannot solve these problems one by one in this paper. We will study them in our future work.